\documentclass[aps,floatfix,prd,twocolumn,a4paper,10pt,citeautoscript,
preprintnumbers,superscriptaddress,showpacs,longbibliography]{revtex4-1} 

\usepackage[english]{babel} 	
\usepackage{amsmath}        	
\usepackage{amssymb}        	
\usepackage{bm}					
\usepackage{bbm}            	
\usepackage{empheq}

\usepackage{graphicx}       
\usepackage{graphics}       
\DeclareGraphicsExtensions{.jpg,.pdf} 
\usepackage{hyperref}
\hypersetup{colorlinks=true, pdfstartview=FitV, 
linkcolor=blue, citecolor=blue, urlcolor=blue}

\newcommand{\Equation}[2]{\begin{equation}\label{#1}#2\end{equation}}
\newcommand{\Align}[2]{\begin{align}\label{#1}#2\end{align}}
\newcommand{\SubAlign}[2]{\begin{subequations}\label{#1}\begin{align}#2\end{align}\end{subequations}}

\newcommand{\bs}{\boldsymbol}
\newcommand{\Figref}[1]{Fig.~\ref{#1}}
\newcommand{\Eqref}[1]{\eqref{#1}}
\newcommand{\Exp}[1]{\text{e}^{#1}}
\renewcommand\Re{\mathrm{Re}}

\newcommand{\Grad}{{\bs\nabla}}

\newcommand{\A}{{\bs A}}
\newcommand{\B}{{\bs B}}
\newcommand{\D}{{\bs \Pi}}

\newcommand{\F}{\mathcal{F}}

\begin{document}
\title{Change of the vortex core structure in two-band superconductors 
at impurity-scattering-driven \texorpdfstring{$s_\pm/s_{++}$}{s+-/s++} 
crossover
}
\author{Julien~Garaud}
\affiliation{Department of Physics,  
KTH-Royal Institute of Technology, Stockholm, SE-10691 Sweden}
\author{Mihail~Silaev}	
\affiliation{Department of Physics and Nanoscience Center, 
University of Jyv\"askyl\"a, P.O. Box 35 (YFL), FI-40014 
University of Jyv\"askyl\"a, Finland}
\author{Egor~Babaev} 
\affiliation{Department of Physics,
KTH-Royal Institute of Technology, Stockholm, SE-10691 Sweden}

\begin{abstract}

We report a nontrivial transition in the core structure of vortices in 
two-band superconductors as a function of interband impurity scattering. 
We demonstrate that, in addition to singular zeros of the order parameter, 
the vortices there can acquire a circular nodal line around the singular 
point in one of the superconducting components. 
It results in the formation of the peculiar ``moat"-like profile in one of the 
superconducting gaps. The moat-core vortices occur generically in the vicinity 
of the impurity-induced crossover between $s_{\pm}$ and $s_{++}$ states. 
  
\end{abstract}

\date{\today}
\maketitle

\begin{figure}[!htb]
\hbox to \linewidth{ \hss
\includegraphics[width=0.99\linewidth]{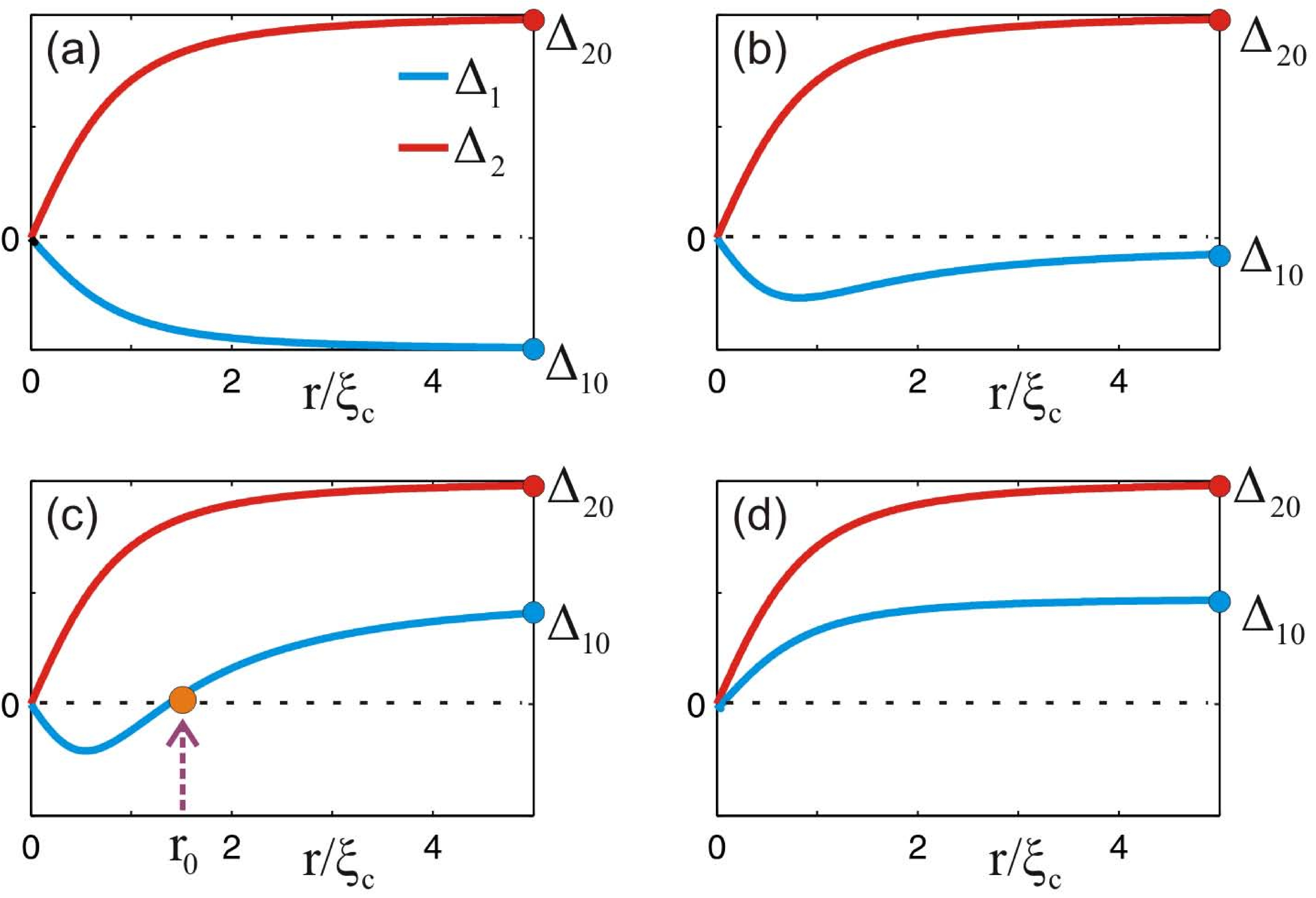}
\hss}
\caption{
(Color online) -- 
Schematic picture illustrating the evolution of gap function profiles 
$\Delta_{1,2}(r)$ near the vortex cores, in two-band superconductors 
when the bulk state undergoes the $s_{\pm}/s_{++}$ crossover. 
Panels (a) and (d) display the usual vortex profiles in the $s_{\pm}$ and 
$s_{++}$ phases respectively. Panel (b) shows a vortex with overshooting 
non-monotonic behavior of the subdominant component $\Delta_1(r)$, while 
panel (c) displays the moat-core vortex in the $s_{++}$ phase with the node 
$\Delta_1(r_0)=0$.
}
\label{Fig:Schematic}
\end{figure}

Singularities that typically occur in quantum vortices are pointlike: 
i.e. in two dimensions, the modulus of the complex order parameter 
(the density of superconducting electrons) vanishes at some point 
in the vortex core \cite{onsager1949statistical,Feynman:55,Abrikosov:57}. 
We consider qualitatively different vortex structures in a rather generic, 
and microscopically simple model of a two-band superconductor with impurities.  
In such a system vortices can have a circular nodal line where the 
superconducting gap function in one of the bands vanishes. In three 
dimensions it extends to a cylindrical nodal surface surrounding the 
vortex line. We introduce the name ``moat-core"-vortex to distinguish 
such an exotic structure, shown schematically in \Figref{Fig:Schematic}(b,c), 
from the usual two-components vortices with monotonic gap profiles 
[\Figref{Fig:Schematic}(a,d)]. 
  
Two-band superconductors where the pairing is generated by interband 
electron-electron repulsion \cite{Hirschfeld.Korshunov.ea:11}, tend 
to form the so-called $s_\pm$ superconducting state with a sign change 
between the gap functions in different bands \cite{Mazin.Singh.ea:08,
Chubukov.Efremov.ea:08} $\Delta_1$ and $\Delta_2$. Namely, there is a $\pi$ 
relative phase between the components $|\Delta_j|e^{i\theta_j}$ of order 
parameter for the band index $j=1,2$. Thus, in contrast to the $s_{++}$ state 
where the ground-state phase difference $\theta_{12}\equiv \theta_{2}-\theta_{1}$ 
is zero, the $s_\pm$ state has $\theta_{12}=\pi$.
Increasing disorder in dirty two-band superconductors rather generically 
leads to a crossover from the $s_{\pm}$ to the $s_{++}$ state. 

For the simplest two-band weak-coupling theory, the crossover can be of two 
types. The first is a direct one \cite{Efremov.Korshunov.ea:11} involving a 
continuous sign change in one of the gap functions, e.g., $\Delta_1$. 
Hereafter we call $\Delta_1$ the subdominant component, since near the 
critical temperature $T_c$, it can be considered as induced by the stronger 
gap $\Delta_2$ due to the Cooper pair interband tunneling. 
The subdominant gap function amplitude vanishes at the crossover line, while 
the leading component $\Delta_2$ remains nonzero. The second scenario involves 
the intermediate time-reversal symmetry breaking $s+is$ state 
\cite{Stanev.Koshelev:14}, when both gap functions $\Delta_{1,2}$ are finite 
but acquire a nontrivial phase difference $\theta_{12}\neq\pi n$. Quantitative 
study has shown that in the second scenario, the intermediate $s+is$ state 
occupies a vanishingly small region of the phase diagram \cite{Silaev.Garaud.ea:17} 
(see also note 
\footnote{ 
It should be noted that this statement applies only to the dirty two-band case, 
in three-band case, the $s+is$ state can occupy much larger fraction of the 
phase diagram \cite{Stanev.Tesanovic:10,Maiti.Chubukov:13,Boeker.Volkov.ea:17}. 
Its experimental observation was recently reported in \cite{Grinenko.Materne.ea:17}.
}). \nocite{Stanev.Tesanovic:10,Maiti.Chubukov:13,Boeker.Volkov.ea:17,
Grinenko.Materne.ea:17}
At the same time, the signature of the $s_{\pm}/s_{++}$ crossover has recently 
been experimentally observed in the superconducting compound from the 
iron-pnictide family with controlled disorder \cite{Schilling2016}.

Here we consider vortex solutions near the $s_{\pm}/s_{++}$ crossover line, 
and demonstrate the formation of moat-core vortices featuring a non-monotonic 
order parameter distribution, and a circular (or cylindrical) nodal line where 
$\Delta_{1}(r_0)=0$. We calculate superconducting ground states and vortex 
structures within the weak-coupling model of two-band superconductors with a 
high concentration of impurities. Such system can be described by two coupled 
Usadel equations with interband impurity scattering terms \cite{Gurevich:03}:
\Align{Eq:Usadel}{
 \omega_n f_i &= \frac{D_i}{2} \big(g_i {\bm \Pi}^2 f_i - f_i \nabla^2 g_i\big) 
 	+  \Delta_i g_i		\nonumber\\
 	&~~~+ \sum_{j\neq i}\gamma_{ij} ( g_if_j - g_jf_i) \,.
}
Here $\omega_n =( 2n+1)\pi T $, $n\in\mathbb{Z}$ are the fermionic Matsubara 
frequencies, and $T$ the temperature. $D_i$ are the electron diffusivities, and 
$\gamma_{ij}$ the interband scattering rates. Propagators in each band obey the 
normalization condition $|f_i|^2 + g_i^2 =1$, where the quasiclassical propagators 
$f_i$ and $g_i$ are, respectively, the anomalous and normal Green's functions. 
The gap functions are determined by the self-consistency equations
\Equation{Eq:SelfConsistency}{
 \Delta_i =2\pi T  \sum_{n=0}^{N_d} 
 \sum_{j} \lambda_{ij} f_{j} (\omega_n)\,,
}
for the Green's functions that satisfy Eq.\Eqref{Eq:Usadel}. Here  
$N_d=\Omega_d/(2\pi T)$ is the summation cutoff at Debye frequency 
$\Omega_d$. The diagonal elements $\lambda_{ii}$ of the coupling matrix 
$\hat{\lambda}$ in the self-consistency equation \Eqref{Eq:SelfConsistency}, 
describe the intraband pairing. The interband interaction is determined by 
the off-diagonal terms $\lambda_{ij}$ ($j\neq i$) which can be either 
positive or negative.

An expansion in small $|\Delta_j| \ll T_c$ and their gradients gives the 
Ginzburg-Landau (GL) model: 
\SubAlign{Eq:FreeEnergy}{
\frac{\F}{\F_0} =&
\sum_{j=1}^2\Big\{
 	\frac{k_{jj}}{2}\left|\D\Delta_j \right|^2
 	+a_{jj}|\Delta_j|^2+\frac{b_{jj}}{2}|\Delta_j|^4\Big\}  
 	\label{Eq:FreeEnergy:Self}	\\
   &+\frac{k_{12}}{2}
	\Big((\D\Delta_1)^*\D\Delta_2+(\D\Delta_2)^*\D\Delta_1 \Big)
	\label{Eq:FreeEnergy:Mixed}	\\
   &+2\left(a_{12}+c_{11}|\Delta_1|^2+c_{22}|\Delta_2|^2\right)
   \Re\big(\Delta_1^*\Delta_2\big)
	\label{Eq:FreeEnergy:Interaction1}	\\
   &+\left(b_{12}+c_{12}\cos2\theta_{12}\right)|\Delta_1|^2|\Delta_2|^2    
   +\frac{\B^2}{2}
	\label{Eq:FreeEnergy:Interaction2}	\,.
}
The two gaps in the different bands are electromagnetically coupled by the 
vector potential $\A$ of the magnetic field $\B=\bs\nabla\times\A$, through 
the covariant derivative $\D\equiv\Grad+iq\A$ where $q$ is the electromagnetic 
coupling constant that parametrizes the magnetic field penetration depth. 
The two components are also directly coupled via potential 
terms in \Eqref{Eq:FreeEnergy:Interaction1} and \Eqref{Eq:FreeEnergy:Interaction2}, 
and via the mixed-gradient term \Eqref{Eq:FreeEnergy:Mixed}.
The coefficients of the Ginzburg-Landau functional $a_{ij}$, $b_{ij}$, $c_{ij}$, 
and $k_{ij}$ can be calculated microscopically.  We list here only the 
expressions for the gradient terms, which  are crucial for the correct 
determination of the  transition in vortex structure:
\SubAlign{Eq:GLparameters:K}{
 k_{ii} &= 2\pi T N_i \sum_{n=0}^{N_d} 
 \frac{ D_i (\omega_n + \gamma_{ji})^2 + \gamma_{ij}\gamma_{ji} D_j }
 {\omega^2_n(\omega_n+\gamma_{ij}+\gamma_{ji})^2} \\
 k_{ij} &= 2\pi T N_i\gamma_{ij} \sum_{n=0}^{N_d} 
 \frac{ D_i (\omega_n + \gamma_{ji}) + D_j (\omega_n + \gamma_{ij}) }
 {\omega^2_n(\omega_n+\gamma_{ij}+\gamma_{ji})^2}	\,,
}
with $j\neq i$, and $N_i=\lambda_{ji}/(\lambda_{12}+\lambda_{21})$ are the 
partial densities of states. Note that the regimes considered below are with 
symmetric interband coupling, thus implying that $N_2/N_1=1$.
The coefficients given by Eq.~\Eqref{Eq:GLparameters:K} satisfy the condition 
$k_{11}k_{22}-k_{12}^2 >0$ yielding the positively defined gradient energy in 
Eq.~\Eqref{Eq:FreeEnergy} for the entire range of diffusivities $D_{1,2}$ and 
effective interband scattering rate $\Gamma = \gamma_{12}/N_2 = \gamma_{21}/N_1$.
The previously reported expressions \cite{Gurevich:03} for $k_{ij}$ violate 
this condition and therefore in general they can be used only for the 
infinitesimally small values of $\Gamma$. 
For the calculations we use dimensionless variables, normalizing the gaps 
by $T_c$, the lengths by $\xi_0 = \sqrt{D_1/T_{c}}$, the magnetic field by 
$B_0 = T_c \sqrt{4\pi \nu_1}$, and the free energy ${\F_0} =B_0^2/4\pi $, 
where $\nu_1$ is the density of states in the first band. The electromagnetic 
coupling constant is $q = 2\pi B_0 \xi_0^2/\Phi_0$. In these units, the London 
penetration length $\lambda_L$ is given by 
$\lambda^{-2}_L=q^2(k_{ii}\Delta_{i0}^2+2 k_{12}\Delta_{10}\Delta_{20})$,
where $\Delta_{i0}$ is the bulk value of the dimensionless gap.

The bulk phase diagram given by this model was calculated both at the 
Ginzburg-Landau level and verified against the numerical solution of the full 
Usadel theory in \cite{Silaev.Garaud.ea:17}. For temperatures rather close to 
$T_c$ it typically displays a direct $s_\pm/s_{++}$ crossover line which is 
rather featureless with respect to thermodynamic signatures \cite{Silaev.Garaud.ea:17}. 
Below, we demonstrate that there is nevertheless a transition in vortex core 
structure across that line. This could have a number of consequences for behavior 
of the system in external magnetic fields.

\begin{figure*}[!htb]
\hbox to \linewidth{ \hss
\includegraphics[width=.925\linewidth]{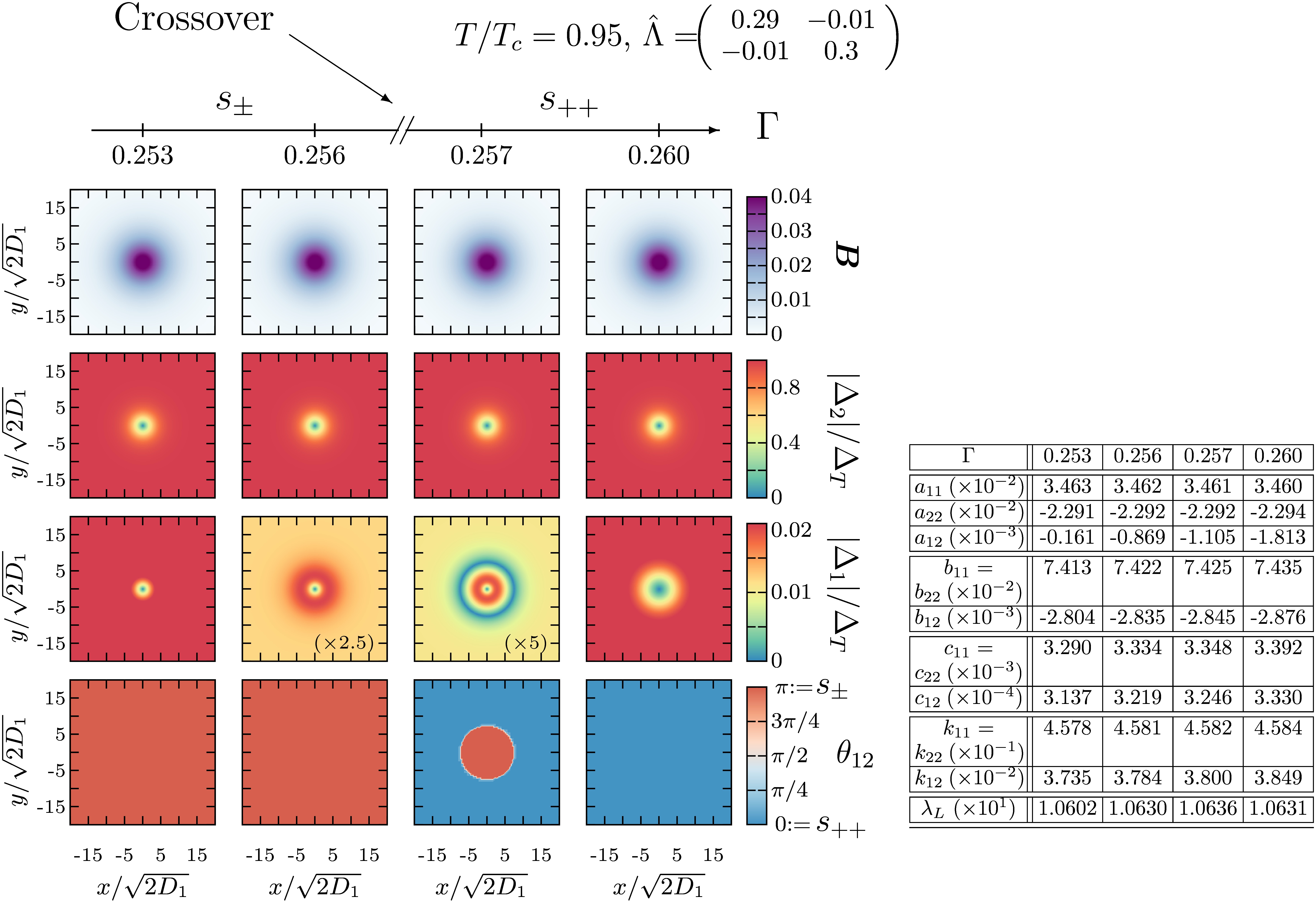}
\hss}
\caption{
(Color online) -- 
Transition in vortex solutions in the vicinity of the 
impurity induced crossover line of a two-band superconductor with nearly 
degenerate bands and weak repulsive interband pairing interaction 
($\lambda_{11}=0.29$, $\lambda_{22}=0.3$, and $\lambda_{12}=\lambda_{21}=-0.01$), 
and with equal electron diffusivities ($D_2/D_1=1$). The temperature is 
$T/T_c=0.95$, $q=0.25$, and tuning the strength of the effective interband 
impurity scattering drives the system from bulk $s_\pm$  to bulk $s_\pm$ $s_{++}$. 
The table indicates the calculated values of the London penetration depth 
$\Lambda_L$ and of the coefficients of the Ginzburg-Landau free energy.
The different lines respectively display the magnetic field $\B$, the larger gap 
($\Delta_2$), and the smaller gap ($\Delta_1$). The last line shows the relative 
phase $\theta_{12}$ that specifies whether the superconducting ground state is 
$s_{++}$ or $s_\pm$. The third column shows a vortex solution that has a point 
like and a ``moat"-like zero in $\Delta_1$. Note this is a close-up view of the 
vortex core which is actually calculated in larger grids.
}
\label{Fig:Vortices}
\end{figure*}

To investigate the properties of single vortex solutions, the physical degrees 
of freedom $\Delta_{1,2}$ and $\A$ are discretized within a finite-element 
formulation \cite{Hecht:12}, and the Ginzburg-Landau free energy \Eqref{Eq:FreeEnergy} 
is minimized using a non-linear conjugate gradient algorithm. Given an initial 
guess where both components have the same phase winding (at large distances 
$\Delta_i\propto\Exp{i\theta}$ and $\theta$ is the polar angle relative to 
the vortex center), the minimization procedure leads, after convergence of 
the algorithm, to a vortex configuration
\cite{
[{Being in zero external field, the vortex is created only by the initial phase 
winding configuration. For further details on the numerical methods employed here, 
see for example related discussion in: }]
[{}] Garaud.Babaev.ea:16}. 
Figure \ref{Fig:Vortices} shows the numerically calculated single vortex solutions 
in the vicinity of the impurity-induced crossover, in the case of a two-band 
superconductor with nearly degenerate bands and weak repulsive interband pairing 
interaction.
There is a transition in the vortex profiles of the subdominant component 
$\Delta_1$ when approaching the $s_\pm/s_{++}$ crossover line. First, we observe 
that on the $s_\pm$ side of the crossover, $\Delta_1(r)$ distribution exhibits 
a strong increase near the core, strongly overshooting its ground-state value 
$\Delta_{10}$ which is retained asymptotically at $r\to \infty$. 
We note that a small density overshoot effect was also obtained in the two-band 
model with ballistic and diffusive bands \cite{Tanaka.Eschrig.ea:07}.
Here we point out that in the vicinity of $s_{++}/s_{\pm}$ crossover, the 
near-core overshoot can be very large, reaching about 100\% of the subdominant 
ground-state amplitude $\Delta_{10}$ [see the examples in \Figref{Fig:Vortices})]. 
The effect should be present rather generically in the presence of the interband 
impurity scattering since it originates from the mixed-gradient term in 
Eq.~\Eqref{Eq:FreeEnergyMixedGrad} that tends to become negative. 
Correspondingly, in the presence of strong superconducting currents it becomes 
under certain conditions energetically beneficial to increase density.

We find that near the $s_{\pm}/s_{++}$ crossover there is a rather generic 
effect of formation of the circular nodal lines of the subdominant component 
$\Delta_{1}=0$. The nodal lines exist in addition to the usual point 
singularities at the vortex center. In that regime, the ground state is 
$s_{++}$ so that the interband phase difference disappears far from the vortex 
center $\theta_{12}\to 0$.
However, due to the competition between gradient and Josephson terms, it 
is more favorable to achieve a $\theta_{12}=\pi$ ($s_{\pm}$ state), in the 
vicinity of the core singularity. The transition between the localized ``core" 
states with $\theta_{12}=\pi$ and the asymptotic states $\theta_{12}=0$ is 
realized by nullifying the subdominant gap $\Delta_1(r_0)=0$ at a given 
distance $r_0$ from the center, when the Josephson energy term wins over 
the gradient one. The effect should also be generic for a wide range of models 
with this structure of the gradient terms competing with the inter-component 
Josephson coupling 
\cite{
[{Here we focus on the role of impurity-induced gradient terms. In fact the 
effect should more general: compare with the phenomenological discussion of 
vortex structure in three-component models without mixed-gradient terms but 
with frustrated inter-band coupling in: }]
[{}] Carlstrom.Garaud.ea:11a}. 
We discuss below that the effect should be stronger at lower temperatures 
and is underestimated by the GL model.

The tendency for the formation of a localized $s_{\pm}$ state inside the vortex 
core can be qualitatively understood by analyzing the functional 
\Eqref{Eq:FreeEnergy}. The structure of axially symmetric vortices is given 
by the ansatz for the order parameter components 
$\Delta_j (\bm r)= \tilde\Delta_j (r) e^{i\theta}$, where $\tilde\Delta_j (r)$ 
are the real-valued profiles of the order parameter components and the polar 
coordinates $r,\theta$ are determined relative to the vortex center. In this 
case the GL energy contribution from the mixed-gradient term can be written 
as follows 
\Align{Eq:FreeEnergyMixedGrad}{ 
F_G &\equiv \frac{k_{12}}{2}
 \Big((\D\Delta_1)^*\D\Delta_2+c.c \Big)  \\ \nonumber
 & = k_{12} \left( \nabla_r \tilde\Delta_1 \nabla_r \tilde\Delta_2 + 
 r^{-2}\tilde\Delta_1\tilde\Delta_2 \right) ,
}
where the vector potential contribution is neglected since it is small 
inside the vortex core. This term describes the interaction between the 
order parameter components which is qualitatively similar to the interband 
Josephson energy contribution in Eq.~\Eqref{Eq:FreeEnergy}:  
\Equation{Eq:FreeEnergyJos}{
 F_J \equiv 2\left(a_{12}+c_{11}|\tilde\Delta_1|^2+c_{22}|\tilde\Delta_2|^2\right)
 \tilde\Delta_1\tilde\Delta_2\,.
}
In the bulk phase, where the gradient energy is zero $F_G=0$, the phase locking 
corresponds to the $s_{++}/s_{\pm}$ state depending on the sign of the effective 
Josephson coupling $J= a_{12}+c_{11}|\tilde\Delta_1|^2+c_{22}|\tilde\Delta_2|^2$. 
The crossover line can be defined parametrically in the $\Gamma$, $T$ plane 
as $J (\Gamma, T) =0$.
In spatially non-homogeneous states, e.g., in the presence of vortices, the 
relative sign of the gap functions $\tilde\Delta_{1,2}$ is determined by the 
local interplay of two phase-locking energies $F_G$ and $F_J$. 
 
In the vortex cores, the order parameter profiles can be approximated by linear 
dependencies $\tilde{\Delta}_{j} (r) \approx r d\tilde{\Delta}_{j}/dr$, thus 
yielding $F_G \approx k_{12} ( d\tilde{\Delta}_{1}/dr )(d\tilde{\Delta}_{2}/dr)$. 
There, since the mixed-gradient coefficient is always positive $k_{12}>0$, the 
energy $F_G$ favors the opposite signs of the order parameter slopes, e.g., 
$ d\tilde{\Delta}_{2}/dr  >0 $ and $d\tilde{\Delta}_{1}/dr <0$. That leads to 
the opposite signs of gap functions near the vortex center $\tilde{\Delta}_{2}>0$ 
and $\tilde{\Delta}_{1}<0$. This tendency competes with that favored by the 
Josephson energy if $J<0$, corresponding to the bulk $s_{++}$ phase when the 
gaps have the same signs far from the core. Therefore, provided that the gradient 
energy dominates close to the vortex center($|F_G| > |F_J|$), one can expect the 
non-monotonic distribution for the component  $\Delta_1(r)$, crossing zero at 
some finite distance $r=r_0$ determined by the competition of $F_G$ and $F_J$.
In the two-dimensional plane perpendicular to the vortex line, such zero points of 
$\Delta_1(r_0)=0$ form the circular nodal line around the singular point at the 
vortex center $r=0$.

The scenario discussed above is actually generic for any two-band $s_{++}$ 
superconductor with interband impurity scattering. It can be shown that the 
effect should be stronger away from the superconducting phase transition. 
For a system that breaks only a single symmetry, at the mean-field level only 
one (critical) mode survives in the limit $\tau \equiv (1-T/T_c) \to +0$. 
In general, the critical mode corresponds to a certain linear combination of 
the gap function fields ${\Delta}_{1,2}$. Even if there are other well-defined 
subdominant modes that are characterized by other coherence lengths, they have 
vanishing amplitude when $\tau$ is much smaller than other parameters in the 
problem 
\cite{
[{See detailed discussion of behavior of dominant and subdominant modes in 
multiband superconductors in the limit $\tau \equiv (1-T/T_c) \to +0$ in: }]
[{}] Silaev.Babaev:12}. 
In the limit $\tau \to 0$, the energy contributions can be estimated by 
retaining only the contribution from the dominant mode, that is, 
$ |d\tilde{\Delta}_{i}/dr | \propto |\Delta_{i0}|/\xi_c(T) $, where 
$\xi_{c}(T)\propto 1/\sqrt{\tau} $ is the critical coherence length. Hence the 
mixed-gradient energy  $|F_G|\propto k_{12} |\Delta_{10}\Delta_{20}|/\xi_c^2(T)$ 
should be compared to the Josephson energy $F_J\propto J \Delta_{10}\Delta_{20}$. 
One can see that the condition of the vortex transition $|F_G| > F_J$ is satisfied 
only provided that the coupling is small enough, $|J|\ll k_{12}/\xi_c^2(T)$, which 
certainly does not hold near the critical temperature in the limit $\tau\to 0$ 
when $\xi_c(T)\to \infty$. However, one can expect that inside the vortex core 
the gradient energy always dominates in the vicinity of the impurity-driven 
$s_{\pm}/s_{++}$ crossover where the effective Josephson coupling disappears, 
$J (\Gamma, T) = 0$. This argument heuristically explains the  numerically 
found moat-core vortex structures shown in \Figref{Fig:Vortices}.

The existence of exotic moat-core vortices does not depend on specific 
values of the pairing coefficients. Indeed, we found such solutions for all 
the different $\hat{\Lambda}$ we investigated. Based on the above qualitative 
argument, one can conclude that these vortex structures inevitably appear 
sufficiently close to the crossover line. Moreover, we find that typically, 
the region of moat-core vortices in the $\Gamma$, $T$ phase diagram tends to 
become larger with the increased ratio of diffusion coefficients $D_{2}/D_1$. 
This effect can be explained by the softening of the order parameter in the 
subdominant band which facilitates the formation of additional zeros in the 
$\Delta_1(r)$ gap distribution.

In conclusion, we have shown that there is a vortex structure transition 
across the $s_\pm/s_{++}$ crossover line driven by the impurity scattering, 
in two-band superconductors. 
On the $s_{\pm}$ side of this crossover, vortices have a strong overshooting 
in the distribution of the subdominant component of the order parameter. On 
the other side, there are moat-core vortices with an $s_\pm$ phase inclusion 
in the cores, separated from the bulk $s_{++}$ phase by circular nodal lines. 
This raises a number of interesting questions. First, it should be interesting 
to investigate the electronic structure of the moat-core vortices. Second, 
this system for the parameters close to the $s_\pm/s_{++}$ crossover should 
have a non-trivial behavior in the external magnetic field. Indeed, in contrast 
to the zero-field picture of a sharp crossover, the lattice and liquids of 
moat-core vortices represent a macroscopic phase separation or  
mircoemulsionlike $s_\pm$ inclusions inside the $s_{++}$ state. As the 
vortex density rises in increasing field, there should also be a field-induced 
crossover from $s_{++}$ to the $s_\pm$. This can be resolved in local 
phase-sensitive probes \cite{hirschfeld2015robust}.

\begin{acknowledgments}
We thank Johan Carlstr\"om for many discussions.
The work was supported by the Swedish Research Council Grants
No. 642-2013-7837 and VR2016-06122 and Goran Gustafsson Foundation 
for Research in Natural Sciences and Medicine. 
M.S. was supported by the Academy of Finland.
The computations were performed on resources 
provided by the Swedish National Infrastructure for Computing 
(SNIC) at National Supercomputer Center at Link\"oping, Sweden.
\end{acknowledgments}


%

\end{document}